\documentclass[traditabstract]{aa}

\begin{document}
\title{Comment on ``Cosmic radio dipole from NVSS and WENSS''}
\titlerunning{Comment on cosmic radio dipole}
\author{Ashok K. Singal}
\institute{Astronomy and Astrophysics Division, Physical Research Laboratory,
Navrangpura, Ahmedabad - 380 009, India\thanks{E-mail:asingal@prl.res.in}}
\authorrunning{A. K. Singal}

\date{}

\abstract{We show that a recent purported correction to the effects of Doppler boosting of 
flux density in an erstwhile published formula for calculating the observer's motion from 
the cosmic radio dipole in sky brightness 
is erroneous. The thereby computed downward correction in the estimated magnitude for the observer's motion  
needs to be scrapped and the results derived therefore need to be reverted back to their erstwhile values.}

\keywords{galaxies: active --- galaxies: statistics --- Local Group ---  cosmic background radiation
--- cosmological parameters --- large-scale structure of universe}
\maketitle
Recently Rubart \& Schwarz (2013, Section 6) claimed a missing factor in an earlier formula given by Singal (2011) 
for calculating the dipole in the radio sky brightness arising due to the observer's peculiar motion. 
Singal (2011) had derived the formula for the dipole resulting from the stellar aberration and Doppler boosting 
of discrete radio sources, briefly in the following manner.
 
An observer moving with a non-relativistic velocity $v$, will find sources in the forward direction brighter 
by a factor  $\delta^{1+\alpha}$, due to Doppler boosting, where $\delta=1+(v/c)\cos\theta$ is the 
Doppler factor, and $\alpha$ ($\approx 0.8$) is the spectral index defined by $S \propto \nu^{-\alpha}$.
With the integral source counts of extragalactic radio source population following a power law  
$N(>S)\propto S^{-x}$ ($x \sim 1$), because of the increased flux density due to Doppler boosting,  
the observed number of sources in any given flux-density range will therefore change by a factor $\propto \delta^{x(1+\alpha)}$. 
In addition, due to the stellar aberration, the number density will be different by a factor $\delta^2$.  
Thus as a combined effect of Doppler boosting and the stellar aberration, the observed sky brightness 
will vary as $\propto \delta^{2+x(1+\alpha)}$  (Ellis \& Baldwin 1984; Crawford 2009) 
which to a first order can be written as $1+{\cal D}\cos\theta$, 
a dipole anisotropy over the sky with amplitude ${\cal D}=[2+x(1+\alpha)](v/c)$. 

If $\theta_i$ is the polar angle of the $i^{th}$ source of observed flux density $S_i$ with respect to the observer's peculiar 
motion, then writing  ${\Delta {\cal F}}=\Sigma S_i\: \cos \theta_i$ and ${\cal F}=\Sigma S_i\: |\cos \theta_i|$ 
and converting the summation into integration over the sphere, we get for the dipole magnitude, 
\begin{eqnarray}
\frac {\Delta {\cal F}}{{\cal F}}=k\: \frac {\int^{\pi}_{0}(1+{\cal D}\cos\theta)\cos \theta\: \sin \theta\:{\rm d}\theta}
{2\int^{\pi/2}_{0}\cos \theta \:\sin \theta\:{\rm d}\theta}=\frac {2k {\cal D}}{3}.
\end{eqnarray}
The formula is equally valid for samples with finite upper and lower flux-density limits. Here $k$ is a constant 
of the order of unity ($k=1$ for a sky fully covered by the sample) and as such may need to be determined numerically for 
individual cases when there are finite gaps in the sky coverage. The  above formula was used to calculate dipole ${\cal D}$ 
from the estimates of ${\Delta {\cal F}}/{{\cal F}}$ for different flux-density bins in the NVSS survey 
that contains a total of about 1.8 million radio sources across the sky.

Rubart \& Schwarz (2013) argued that since the Doppler effect alters the observed fluxes, this change in 
flux density is an additional factor along with the change in number counts at a given flux density because of the power law 
$n(S) \propto S^{-\tilde x}$. Accordingly, combining the Doppler boosting and stellar aberration and keeping only first-order terms, 
they got for the differential number counts, 
\begin{equation}
 \frac{\mathrm{d}^2N}{\mathrm{d} \Omega \mathrm{d}S} 
 =n_0(S_0)[1+(2+\tilde x (1+\alpha))\frac{v}{c}\cos \theta],
\end{equation}
Then the dipole could be obtained from the summation, 
\begin{equation}
 \int_{4\pi}  \mathrm{d}\Omega 
 \int_{S_\mathrm{min}}^{S_\mathrm{max}} \mathrm{d}S
 \frac{\mathrm{d}^2N}{\mathrm{d} \Omega \mathrm{d}S}  S \cos\theta.
\end{equation}

From (3) they got essentially the same formula as (1), derived by Singal (2011), except that $x$ was replaced by $\tilde x$. 
Of course it goes without saying 
that any second-order effects that might arise because of the number counts not being a simple power-law or the  
changes, if any, in the spectral index over the flux-density range used will have to be considered. 
But the main contention of Rubart \& Schwarz (2013) here was that $\tilde x = 1 + x$ (or larger to also account for the 
steepening of the spectral index at high fluxes), thus making ${\cal D}=[2+(x+1)(1+\alpha)](v/c)$ in Singal's formulation. 
This increases the expectation value  $\langle{\cal D}_{\mathrm{obs}} \rangle$ of the observed dipole by,
\begin{equation}
 \frac{\langle {\cal D}_{\mathrm{obs}}\rangle|_{\tilde x=x+1}}{ \langle {\cal D}_{\mathrm{obs}} \rangle|_{x}} 
 \stackrel{>}{_{\sim}} 1.4,
\end{equation}
for $x\approx 1$. Rubart \& Schwarz (2013) assert that the results of the flux-density weighted number counts in Singal (2011) 
are accordingly higher at least by a factor of $1.4$, 
implying that results of the observer's motion inferred from the dipole in the sky brightness,  
as estimated in Singal (2011), should be reduced by a factor of $1.4$, basically 
because of the fact that the appropriate exponent of the differential number count is given by $\tilde{x} = x+1$. 

Actually the above argument by Rubart \& Schwarz (2013) is erroneous as the contribution to the 
observed sky brightness at a given flux-density level S comes from sources whose rest-frame 
flux density is $S/\delta^{1+\alpha}$. The flux boosting of individual sources, pointed out by them in the formula gets 
compensated exactly because of the fact that in the rest frame the sources were intrinsically weaker 
by $\delta^{(1+\alpha)}$. With the integral source counts of extragalactic radio source 
population following a power law $N(>S) \propto S^{-x}$, it is only the number of sources at the flux density S that will change  
by a factor $\delta^{x(1+\alpha)}$. The boosting in the flux density of a source 
by factor $\delta^{1+\alpha}$ is already accounted for, and the excess in sky brightness in the forward 
direction in the moving observer's frame is only because there is larger number of sources in each considered flux-density bin. 

The fallacy in the Rubart \& Schwarz (2013) argument can be seen in the following manner too (see also Tiwari et al. 2014). 
While it is true that the differential source counts $\mathrm{d}N/\mathrm{d}S\propto S^{-(x+1)}$ in the rest frame, 
and at a first look it may seem that because of the Doppler boosting of flux density from $S/\delta^{1+\alpha}$ in the rest 
frame to $S$ in a moving observer's frame, the differential counts $\mathrm{d}N/\mathrm{d}S$ in observer's frame will be related 
to that in the rest frame by $\delta^{(x+1)(1+\alpha)}$, but that is not correct. The number of sources 
$\mathrm{d}N/\mathrm{d}S$ d$S$ seen in the moving observer's frame between $S$ and $S$+d$S$ will be the same as that lies  
between $S/\delta^{1+\alpha}$ and $(S$+d$S)/\delta^{1+\alpha}$ in the rest frame, i.e., 
$\mathrm{d}N/\mathrm{d}S$ d$S \propto S^{-(x+1)} \delta^{(x+1)(1+\alpha)}$ d$S /\delta^{1+\alpha}$. 
Therefore $\mathrm{d}N/\mathrm{d}S\propto S^{-(x+1)}\delta^{x(1+\alpha)}$ in the moving observer's frame. 
Even otherwise a power law for the integral source counts implies that the number of sources $N(>S)$ in the moving observer's frame is 
$N(>S/\delta^{1+\alpha})$ in the rest frame because flux density S in the moving observer's frame implies a rest-frame 
flux density $S/\delta^{1+\alpha}$. Therefore in moving observer's frame we have $N(>S)\propto S^{-x}\delta^{x(1+\alpha)}$. 
Differentiating we get $\mathrm{d}N/\mathrm{d}S\propto S^{-(x+1)}\delta^{x(1+\alpha)}$ in the moving observer's frame, as was the 
inference earlier.

Thus we see that the additional factor $\delta^{1+\alpha}$, pointed out by Rubart \& Schwarz (2013) in the 
formula for dipole derived by Singal (2011), is erroneous and therefore the correction suggested to the estimates of the observer's 
motion from the dipole in the sky brightness is superfluous and needs to be dropped.


\begin{thebibliography}{}
\bibitem{} Crawford, F., 2009, ApJ, 692, 887 
\bibitem{} Ellis, G. F. R. \& Baldwin, J. E., 1984, MNRAS, 206, 377
\bibitem{33} Rubart, M. \& Schwarz, D. J., 2013, A\&A, 555, A117
\bibitem{17} Singal, A. K., 2011, ApJ, 742, L23 
\bibitem{34} Tiwari, P., Kothari, R., Naskar, A., Nadkarni-Ghosh, S. \& Jain, P., 2014, Astroparticle Physics 
(in press), arXiv:1307.1947v4
\end{thebibliography}
\end{document}